\documentstyle[manuscript,aps,prd,psfig,tighten]{revtex}
\begin{document}
\draft
\preprint{OKHEP--98-02}

\title{Difficulties with Photonic Searches for Magnetic Monopoles}

\author{Leonard Gamberg,\thanks{Electronic address: gamberg@mail.nhn.ou.edu}
George R. Kalbfleisch,\thanks{Electronic address: grk@mail.nhn.ou.edu} and
Kimball A. Milton\thanks{Electronic address: milton@mail.nhn.ou.edu}}
\address{Department of Physics and Astronomy, University of Oklahoma, 
Norman, OK
73019, USA}

\date{\today}
\maketitle

\begin{abstract}
Recently, there have been proposals that the classic Euler-Heisenberg
Lagrangian together with duality
 could be employed to set limits on magnetic monopoles having
masses less than 1 TeV.  The D0 collaboration at Fermilab has used
such a proposal to set mass limits based on the nonobservation of pairs of
photons each with high transverse momentum.  In this note, we critique the 
underlying theory, by showing that at the quoted limits the cross section 
violates unitarity and is unstable with respect to radiative corrections.
It is proposed that the correct coupling of magnetic monopoles to photons
leads to an effective softening of the interaction,
leading to a much smaller cross section, from which
no significant limit can be obtained from the current experiments.
Previous limits based on virtual monopole loops are similarly criticized.
\end{abstract}

\section{Introduction}
The notion of magnetic charge has intrigued physicists since Dirac \cite{dirac}
showed that it was consistent with quantum mechanics provided a suitable 
quantization condition was satisfied: For a monopole of magnetic charge $g$ in 
the presence of an electric charge $e$, that quantization condition is
(in this paper we use rationalized units)
\begin{equation}
{eg\over4\pi}={n\over2}\hbar c,
\label{quant}
\end{equation}
where $n$ is an integer.  For a pair of dyons, that is, particles carrying
both electric and magnetic charge, the quantization condition is replaced
by \cite{schwinger}
\begin{equation}
{e_1g_2-e_2g_1\over4\pi}={n\over 2}\hbar c,
\label{squant}
\end{equation}
where $(e_1,g_1)$ and $(e_2,g_2)$ are the charges of the two dyons.

With the advent of ``more unified'' non-Abelian theories, classical composite
monopole solutions were discovered \cite{nonabel}.  The mass of these monopoles
would be of the order of the relevant gauge-symmetry breaking scale,
which for grand unified theories is of order
$10^{16}$ GeV or higher.  But there are models where the electroweak symmetry
breaking can give rise to monopoles of mass $\sim 10$ TeV \cite{ewmono}.
 Even the latter are not yet accessible to accelerator experiments, 
 so limits on heavy monopoles depend either on cosmological considerations
\cite{cosmo}, or detection of cosmologically produced (relic) monopoles
impinging upon the earth or moon \cite{relic}.
However, {\it a priori}, there is no reason that Dirac/Schwinger monopoles
or dyons of arbitrary mass might not exist: It is important to set limits
below the 1 TeV scale.

Such an experiment is currently in progress at the University of Oklahoma
\cite{ou}, where we expect to be able to set limits on {\it direct\/} monopole
production at Fermilab up to several hundred GeV.  This will be a substantial
improvement over previous limits \cite{prev}.  But {\it indirect\/} 
searches have
been proposed and carried out as well.  De R\'ujula \cite{DeRujula} proposed 
looking at the three-photon decay of the $Z$ boson, where the process proceeds
through a virtual monopole loop. If we use his formula \cite{DeRujula} for the
branching ratio for the $Z\to3\gamma$ process, compared to the current
experimental upper limit \cite{exp}
for the branching ratio of $10^{-5}$, we can rule out
monopole masses lower than about 400 GeV, rather than the 600 GeV quoted in
Ref.~\cite{DeRujula}.
Similarly, Ginzburg and Panfil \cite{ginz1} and very recently Ginzburg and
Schiller \cite{ginz2} considered the production of two photons with
high transverse momenta by the collision of two photons produced either
from $e^+e^-$ or quark-(anti-)quark collisions. Again the final photons are 
produced through a virtual monopole loop.  Based on this theoretical scheme,
 an experimental limit has 
appeared by the D0 collaboration \cite{d0}, which sets the following bounds
on the monopole mass $M$:
\begin{equation}
{M\over n}>\left\{\begin{array}{cc}
610 \mbox{ GeV}&\mbox{ for } S=0\\
870 \mbox{ GeV}&\mbox{ for } S=1/2\\
1580 \mbox{ GeV}&\mbox{ for } S=1 \end{array}\right.,
\end{equation}
where $S$ is the spin of the monopole.
It is worth noting that a mass limit of 120 GeV for a Dirac monopole has been
set by Graf, Sch\"afer, and Greiner \cite{graf}, based on the monopole
contribution to the vacuum polarization correction to the muon anomalous
magnetic moment. (Actually, we believe that the correct limit, obtained
from the well-known textbook formula \cite{js1} for the $g$-factor correction
due to a massive Dirac particle is 60 GeV.)

The purpose of this paper is to critique the theory of Refs.~\cite{DeRujula},
\cite{ginz1}, \cite{ginz2}, and \cite{graf}. We will show that it is based on 
a naive application of electromagnetic duality; the resulting
cross section cannot be valid because
unitarity is violated for monopole masses as low as the quoted limits, and
the process is subject to enormous, uncontrollable radiative corrections.
It is not correct, in any sense, as Refs. \cite{ginz2} and \cite{d0}
state, that the effective expansion parameter is $g\omega/M$, where $\omega$
is some external photon energy; rather, the factors of $\omega/M$ emerge
kinematically from the requirements of gauge invariance at the one-loop
level.  If, in fact, a correct calculation introduced such 
additional factors of $\omega/M$, arising from the complicated coupling
of magnetic charge to photons, we argue that no limit could be deduced
for monopole masses from the current experiments. It may even be the
case, based on preliminary field-theoretic calculations, that processes
involving the production of real photons vanish.

\section{Duality and the Euler-Heisenberg Lagrangian}

\begin{figure}
\centerline{\psfig{figure=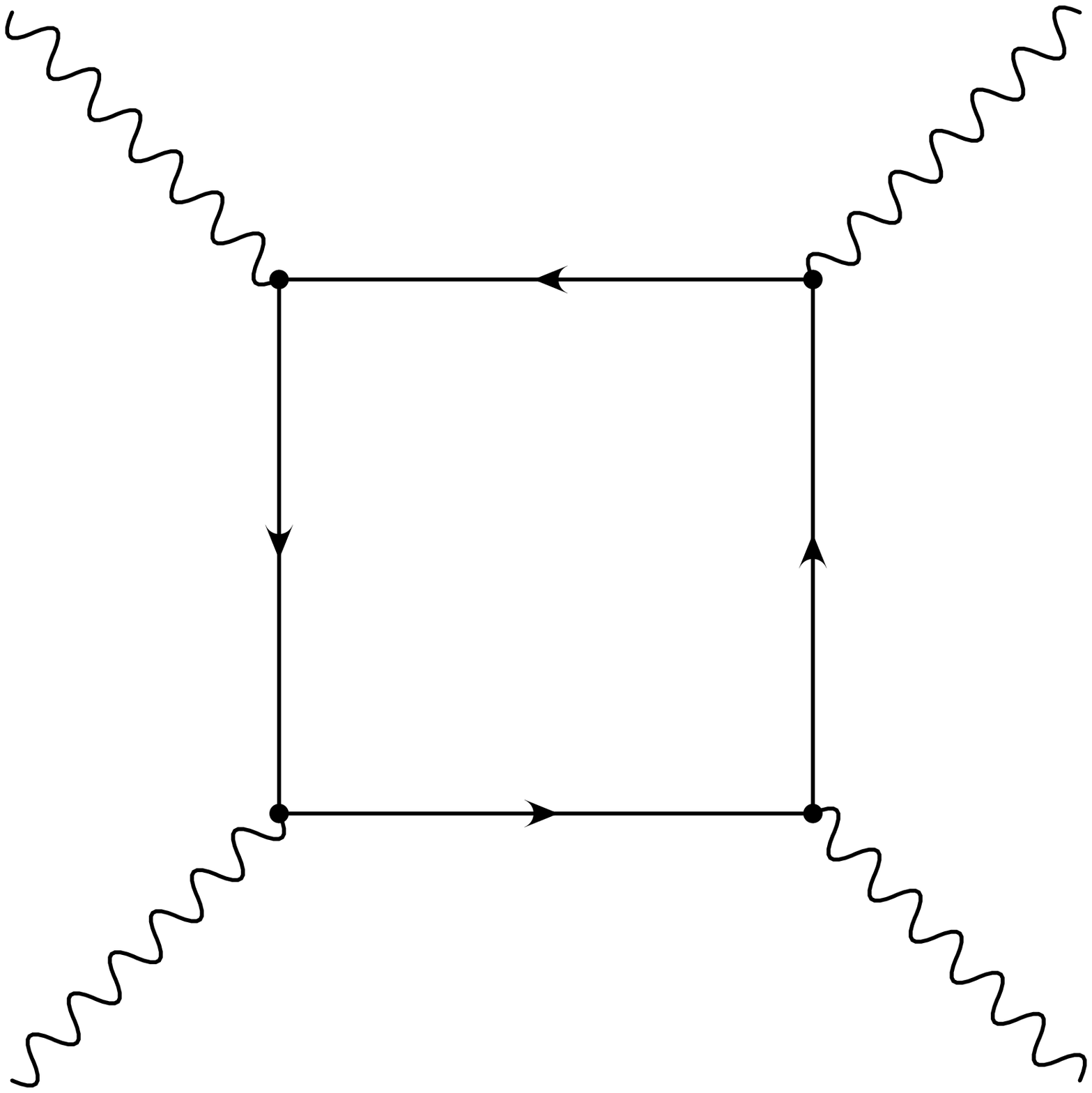,height=3in,width=3in,angle=270}}
\caption{The light-by-light scattering graph for either an electron
or a monopole loop.}
\label{fig1}
\end{figure}

Let us concentrate on the process contemplated in Refs.~\cite{ginz2}
and \cite{d0}, that is
\begin{equation}
\left(\begin{array}{ccc}
qq&\to& qq\\
\bar qq&\to&\bar qq
\end{array}\right)+\gamma\gamma,\quad \gamma\gamma\to\gamma\gamma,
\end{equation}
where the photon scattering process is given by the one-loop light-by-light
scattering graph shown in Fig.\ \ref{fig1}.  If the particle in the loop
is an ordinary electrically charged electron, this process is 
well-known \cite{js,js1,ll}. If, further, the photons involved are of very low
momentum compared the the mass of the electron, then the result may be
simply derived from the well-known Euler-Heisenberg Lagrangian \cite{eh}, which
for a spin 1/2 charged-particle loop in the presence of homogeneous 
electric and magnetic fields is\footnote{We emphasize that 
Eq.~(\ref{ehlagrangian}) is only valid when $\partial_\alpha F_{\mu\nu}=0$.}
\begin{equation}
{\cal L}=-{\cal F}-{1\over8\pi^2}\int_0^\infty {ds\over s^3}e^{-m^2 s}
\left[(es)^2{\cal G}{\mbox{Re}\cosh esX\over\mbox{Im}\cosh esX}-1-{2\over3}
(es)^2{\cal F}\right].
\label{ehlagrangian}
\end{equation}
Here the invariant field strength combinations are
\begin{equation}
{\cal F}={1\over 4}F^2={1\over2}({\bf H}^2-{\bf E}^2),\quad
{\cal G}={1\over 4}F \tilde F={\bf E\cdot H},
\end{equation}
$\tilde F_{\mu\nu}={1\over2}\epsilon_{\mu\nu\alpha\beta}F^{\alpha\beta}$ being
the dual field strength tensor, and the argument of the hyperbolic cosine
in Eq.~(\ref{ehlagrangian}) is given in terms of
\begin{equation}
X=[2({\cal F}+i{\cal G})]^{1/2}=[({\bf H}+i{\bf E})^2]^{1/2}.
\end{equation}
If we pick out those terms quadratic, quartic and 
sextic in the field strengths,
we obtain\footnote{Incidentally, note that the coefficient of the last term is 
36 times larger than that given in Ref.~\cite{DeRujula}.}
\begin{eqnarray}
{\cal L}&=&-{1\over4}F^2+{\alpha^2\over360}{1\over m^4}
[4(F^2)^2+7(F \tilde F)^2]\nonumber\\
&&\mbox{}-{\pi\alpha^3\over630}{1\over m^8}F^2[8(F^2)^2+13 (F {}^*F)^2]+\dots.
\label{ehlag}
\end{eqnarray}
The Lagrangian for a spin-0 and spin-1 charged particle in the loop 
is given by similar formulas which are derived in Ref.~\cite{js,js1}
and (implicitly) in Ref.~\cite{spin1}, respectively.

Given this homogeneous-field effective Lagrangian, it is a simple matter to
derive the cross section for the $\gamma\gamma\to\gamma\gamma$ process in the
low energy limit. (These results can, of course, be directly calculated
from the corresponding one-loop Feynman graph with on-mass-shell photons.
See Refs.~\cite{js1,ll}.)
Explicit results for the differential cross section are given 
by Ref.~\cite{ll}:
\begin{equation}
{d\sigma\over d\Omega}={139\over32400\pi^2}\alpha^4{\omega^6\over m^8}
(3+\cos^2\theta)^2,
\end{equation}
and the total cross section for a spin-1/2 charged particle in the loop 
is\footnote{The numerical coefficient in the total cross section for a spin-0
and spin-1 charged particle in the loop is $119/20250\pi$ and $2751/250\pi$,
respectively.  Numerically the coefficients are $0.00187$, $0.0306$, and
$3.50$ for spin 0, spin 1/2, and spin 1, respectively.}
\begin{equation}
\sigma={973\over10125\pi}\alpha^4{\omega^6\over m^8}.
\label{llcs}
\end{equation}
Here, $\omega$ is the energy of the photon in the center of mass frame,
$s=4\omega^2$. This result is valid provided $\omega/m\ll 1$.
The dependence on $m$ and $\omega$ is evident from the 
Lagrangian (\ref{ehlag}),
the $\omega$ dependence coming from the field strength tensor.
Further note that perturbative quantum corrections are small, 
because they are of relative order $3\alpha\sim10^{-2}$ \cite{dicus}.  
Processes in which
four final-state photons are produced, which may be easily calculated from
the last displayed term in Eq.~(\ref{ehlag}), are even smaller, being
of relative order $\sim\alpha^2 (\omega/m)^8$.  So light-by-light scattering,
which has been indirectly observed through its contribution to the
anomalous magnetic moment of the electron \cite{anmm}, is completely under
control for electron loops.

How is this applicable to photon scattering through a monopole loop? At first
blush this calculation seems formidable.  The interaction of a magnetically
charged particle with a photon involves a ``string,'' that is, an arbitrary
vector function $f_\mu(x-x')$ that satisfies
\begin{equation}
\partial_\mu f^\mu(x-x')=\delta(x-x'),
\end{equation}
which can be realized by a line integral, for example, the semi-infinite
one
\begin{equation}
f_\mu(x)=\int_0^\infty d\xi_\mu\,\delta(x-\xi),
\end{equation}
where the $\xi$ integration follows some path from the origin to infinity.
For the case of a straight line with direction $n_\mu$, this can be
written in the form
\begin{equation}
f_\mu(x)={n_\mu\over i}\int{(dq)\over(2\pi)^4}{e^{iqx}\over n\cdot q
-i\epsilon}.
\label{string}
\end{equation}
The interaction between a magnetic current $J_m^\mu$
 and the electromagnetic field is given by
\begin{equation}
W_{\rm int}=\int (dx)(dx')\tilde F_{\mu\nu}(x')f^\nu(x'-x)J_m^\mu(x),
\label{mmint}
\end{equation}
where the magnetic current must be conserved, $\partial_\mu J_m^\mu=0$.
Here, the string-dependent field strength tensor \cite{dirac,schwinger}
is
\begin{equation}
F_{\mu\nu}(x)=\partial_\mu A_\nu-\partial_\nu A_\mu+\epsilon_{\mu\nu\sigma\tau}
\int(dy)f^\sigma(x-y)J^\tau_m(y).
\end{equation}
The interaction (\ref{mmint}) corresponds to a coupling between electric
and magnetic currents of
\begin{equation}
W^{(eg)}=
-\epsilon_{\mu\nu\sigma\tau}\int(dx)(dx')(dx'')J_e^\mu(x)\partial^\sigma
D_+(x-x')f^\tau(x'-x'')J_m^\nu(x'').
\label{weg}
\end{equation}
From Eqs.~(\ref{string})--(\ref{weg}) 
one obtains the relevant string-dependent 
monopole-photon coupling vertex in momentum space,
\begin{equation}
\Gamma_\mu(q)=ig{\epsilon_{\mu\nu\sigma\tau} n^\nu q^\sigma\gamma^\tau\over
n\cdot q-i\epsilon}.
\label{mmvertex}
\end{equation}

The choice of the string is arbitrary; reorienting the string is a kind
of gauge transformation.  In fact, it is this requirement that leads to
the quantization conditions (\ref{quant}) and (\ref{squant}).  The consistency
of magnetic charge has been demonstrated in quantum mechanics (for
example, see Refs.~\cite{dyondyon} and \cite{dm}), 
but never completely in quantum field theory.\footnote{Arguments have been
given to demonstrate the relativistic invariance of the theory, and the
string independence of the action for classical particle currents
\cite{schwinger}.  See also
Ref.~\cite{brandt}. This consistency is a consequence of the quantization
condition (\ref{quant}) or (\ref{squant}).  We should also bear in mind
Schwinger's warning, in the first reference in Ref.~\cite{schwinger}:
``Relativistic invariance will appear to be violated
in any treatment based on a perturbation expansion.  Field theory is
more than a set of `Feynman's rules.'~'' } The use of the string-dependent
vertex (\ref{mmvertex}) directly is not meaningful.  In this regard,
the ``remedy'' proposed by Deans \cite{deans} and cited as a solution
to the gauge-string dependence of Drell-Yan processes is implausible.

The authors of Refs.~\cite{DeRujula}, \cite{ginz1}, and \cite{ginz2} do not
attempt a calculation of the ``box'' diagram with the 
interaction (\ref{mmint}).
Rather, they (explicitly or implicitly) appeal to duality, that is, the
symmetry that the introduction of magnetic charge brings to Maxwell's 
equations:
\begin{equation}
{\bf E}\to {\bf H}, \quad {\bf H}\to -{\bf E},
\label{duality}
\end{equation}
and similarly for charges and currents.  Thus the argument is that 
for low energy photon processes
it suffices to compute the fermion loop graph in the presence of
zero-energy photons, that is, in the presence of static, constant fields.
The box diagram shown in Fig.~\ref{fig1} with a spin-1/2 
monopole running around the loop in the presence of a homogeneous $\bf E, H$ 
field is then obtained from that analogous process with an electron in the 
loop in the presence of a homogeneous $\bf H, -E$ field, with the substitution 
$e\to g$.  Since the Euler-Heisenberg Lagrangian (\ref{ehlag}) is invariant 
under the substitution (\ref{duality}) on the fields alone, 
this means we obtain the low energy 
cross section $\sigma_{\gamma\gamma\to\gamma\gamma}$ through the monopole
loop from Eq.~(\ref{llcs}) by the substitution $e\to g$, or
\begin{equation}
\alpha\to\alpha_g={137\over 4}n^2,\quad n=1,2,3,\dots.
\label{subs}
\end{equation}

\section{Inconsistency of the Duality Approximation}

It is critical to emphasize that the Euler-Heisenberg Lagrangian is 
an effective
Lagrangian for calculations at the {\it one fermion loop level\/} for low
energy, i.e., $\omega/M\ll1$. It is commonly asserted that the Euler-Heisenberg
Lagrangian is an {\it effective Lagrangian\/} in the sense used in chiral
perturbation theory \cite{weinberg,halter}.  This is not true.  The QED
expansion generates derivative terms which do not arise in the effective
Lagrangian expansion of the Euler-Heisenberg Lagrangian \cite{dicus}.
One can only say that the Euler-Heisenberg Lagrangian is a good approximation
for light-by-light scattering (without monopoles)
at low energy because radiative corrections are down by factors of 
$\alpha$.  However, it becomes unreliable if radiative corrections are
large.

In this regard, both the Ginzburg \cite{ginz1,ginz2} and the De R\'ujula
\cite{DeRujula} articles, particularly Ref.~\cite{ginz2}, are rather misleading
as to the validity of the approximation sketched in the previous section.  
They state that the expansion parameter is not $g$ but $g\omega/M$, $M$ being 
the monopole mass, so that the perturbation expansion may be valid for large 
$g$ if $\omega$ is small enough.  But this is an invalid argument.  
It is only when external photon lines are attached that 
extra factors of $\omega/M$ occur, due to the appearance of the field
strength tensor in the Euler-Heisenberg Lagrangian.  Moreover, the powers
of $g$ and $\omega/M$ are the same only for the $F^4$ process.
The expansion parameter
is $\alpha_g$, which is huge.  Instead of radiative corrections being of the
order of $\alpha$ for the electron-loop process, these corrections will
be of order $\alpha_g$, which implies an uncontrollable sequence of
corrections.  For example, the internal radiative correction to the box
diagram in Fig.~\ref{fig1} have been computed by Ritus \cite{ritus} and by
Reuter, Schmidt, and Schubert \cite{reuter} in QED.  In the $O(\alpha^2)$
term in Eq.~(\ref{ehlag}) the coefficients of the $(F^2)^2$ and the 
$(F\tilde F)^2$ terms are multiplied by $\left(1+{40\over9}{\alpha\over\pi}
+O(\alpha^2)\right)$ and $\left(1+{1315\over252}{\alpha\over\pi}+O(\alpha^2)
\right)$,
respectively.  The corrections become meaningless when we {\it replace\/}
$\alpha\to\alpha_g$.

 This would seem to be a
devastating objection to the results quoted in Ref.~\cite{ginz2} and used
in Ref.~\cite{d0}.  But even if one closes one's eyes to higher order effects,
it seems clear that the mass limits quoted are inconsistent.

If we take the cross section given by Eq.~(\ref{llcs}) and make the
substitution (\ref{subs}), we obtain for the low energy light-by-light
scattering cross section in the presence of a monopole loop
\begin{equation}
\sigma_{\gamma\gamma\to\gamma\gamma}\approx {973\over2592000\pi}
{n^8\over\alpha^4}{\omega^6\over M^8}=4.2\times 10^4 \,n^8{1\over M^2}\left(
\omega\over M\right)^6.
\label{monocs}
\end{equation}
If the cross section were dominated by a single partial wave of angular
momentum $J$, the cross section would be bounded by
\begin{equation}
\sigma\le{\pi(2J+1)\over s}\sim {3\pi\over s},
\end{equation}
if we take $J=1$ as a typical partial wave. Comparing this with the
cross section given in Eq.~(\ref{monocs}), we obtain the following
inequality for the cross section to be consistent with unitarity,
\begin{equation}
{M\over\omega}\gtrsim 3 n.
\label{unitbd}
\end{equation}  
But the limits quoted for the monopole mass are less than this:
\begin{equation}
{M\over n}>870 \mbox{ Gev}, \quad \mbox{spin } 1/2,
\end{equation}
because, at best, a minimum $\langle\omega\rangle\sim 300$ GeV; 
the theory cannot sensibly be applied below a monopole mass of about 1 TeV.
  (Note that changing the
value of $J$ in the unitarity limits has very little effect on the bound
(\ref{unitbd}) since an 8th root is taken: replacing $J$ by 50 reduces
the limit (\ref{unitbd}) only by 50\%.)

Similar remarks can be directed toward the De R\'ujula limits \cite{DeRujula}. 
That author, however, notes the ``perilous use of a perturbative expansion
in $g$.''  However, he fails to use the correct vertex, Eq.~(\ref{mmvertex}),
instead appealing to duality, and even so he admittedly omits
enormous radiative corrections of $O(\alpha_g)$ without any justification
other than what we believe is a specious reference to the use of effective
Lagrangian techniques for these processes.

\section{Proposed Remedies}

Apparently, then, the formal small $\omega$ result obtained from the 
Euler-Heisenberg Lagrangian cannot be valid beyond a photon energy 
$\omega/M\gtrsim0.1$. The reader might ask why one cannot use
duality to convert the monopole coupling with an arbitrary photon to
the ordinary vector coupling.  The answer is that little is thereby
gained, because the coupling of the photon to ordinary charged particles
is then converted into a complicated form analogous to Eq.~(\ref{mmint}).
This point is stated and then ignored in Ref.~\cite{DeRujula} in the
calculation of $Z\to3\gamma$.
There is, in general, no way of avoiding the complication of including
the string.

We are currently undertaking realistic calculations of virtual (monopole
loop) and real (monopole production) magnetic monopole processes \cite{gm}.  
These calculations are, as the reader may infer, somewhat difficult
and involve subtle issues of principle involving the string, and it will
be some time before we have results to present.  Therefore, here we wish
to offer plausible qualitative considerations, which we believe 
suggest bounds that call into question
the results of Ginzburg et al. \cite{ginz1,ginz2}.

Our point is very simple.  The interaction (\ref{mmint}) couples the magnetic
current to the dual field strength.  This corresponds to
the velocity suppression in the interaction 
of magnetic fields with electrically charged particles,
or to the velocity suppression in the interaction 
 of electric fields with magnetically charged
particles, as most simply seen in the magnetic analog of the Lorentz force,
\begin{equation}
{\bf F}=g({\bf B}-{{\bf v}\over c}\times {\bf E}).
\end{equation}
That is, the force between an electric charge $e$ and magnetic charge $g$,
moving with relative velocity $\bf v$ and with relative separation $\bf r$
is
\begin{equation}
{\bf F}=-{eg\over c}{{\bf v\times r}\over4\pi r^3}.
\end{equation}
This velocity suppression is reflected in nonrelativistic calculations.
For example, the energy loss in matter
of a magnetically charge particle is approximately obtained from that
of a particle with charge $Ze$ by the substitution \cite{ce}
\begin{equation}
{Ze\over v}\to{g\over c}.
\end{equation}
And the classical nonrelativistic dyon-dyon scattering cross section near
the forward direction is \cite{dyondyon}
\begin{equation}
{d\sigma\over d\Omega}\approx{1\over(2\mu v)^2}\left[\left(e_1g_2-e_2g_1
\over4\pi c\right)^2+\left(e_1e_2+g_1g_2\over4\pi v\right)^2\right]
{1\over(\theta/2)^4},\quad \theta\ll1,
\end{equation}
the expected generalization of the Rutherford scattering cross section at
small angles.  

Of course, the true structure of the magnetic interaction and the resulting 
scattering cross section is much more complicated.  For example, classical 
electron-monopole or dyon-dyon scattering exhibits rainbows and glories, 
and the quantum scattering exhibits
a complicated oscillatory behavior in the backward direction \cite{dyondyon}.  
These reflect the complexities of the magnetic interaction between 
electrically and magnetically charged particles, which can be represented as a 
kind of angular momentum \cite{dm,gold}.  Nevertheless, for the purpose of
extracting qualitative information, the naive substitution,
\begin{equation}
e\to {v\over c}g,
\end{equation}
seems a reasonable first step.\footnote{This, and the extension of this
idea to virtual processes, leaves aside the troublesome issue
of radiative corrections.  The hope is that an effective Lagrangian
can be found by approximately integrating over the fermions 
which incorporates these effects.}
 Indeed, such a substitution was used in the
proposal \cite{ou} to estimate production rates of monopoles at Fermilab.

The situation is somewhat less clear for the virtual processes
considered here.  Nevertheless, the interaction (\ref{mmint}) suggests
there should, in general, be a softening of the vertex. In the current
absence of a valid calculational scheme, we will merely suggest two
plausible alternatives to the mere replacement procedure adopted in
Refs.~\cite{DeRujula,ginz1,ginz2,graf}.

We first suggest, as seemingly Ref.~\cite{ginz2} does, 
that the approximate effective vertex incorporates 
an additional factor of $\omega/M$.
Thus we propose the following estimate for the $\gamma\gamma$ cross
section in place of Eq.~(\ref{monocs}),
\begin{equation}
\sigma_{\gamma\gamma\to\gamma\gamma}\sim10^4 n^8{1\over M^2}\left(\omega\over
M\right)^{14},
\label{goodcs}
\end{equation}
since there are four suppression factors in the amplitude.
Now a considerably larger value of $\omega$ is consistent with unitarity,
\begin{equation}
{M\over\omega}\gtrsim\sqrt{3n},
\end{equation}
if we take $J=1$ again.  We now must re-examine the $\sigma_{pp\to\gamma
\gamma X}$ cross section.

In the model
given in Ref.~\cite{ginz2}, where the photon energy distribution is
given in terms of the functions $f(y)$, $y=\omega/E$, 
the physical cross section is given by
\begin{equation}
\sigma_{pp\to\gamma\gamma X}=\left(\alpha\over\pi\right)^2\int{dy_1\over y_1}
{dy_2\over y_2}f(y_1)f(y_2)\sigma_{\gamma\gamma\to\gamma\gamma}=
\int dy_1 dy_2{d\sigma\over dy_1\,dy_2},
\end{equation}
where now ({\it cf.} Eq.~(25) of Ref.~\cite{ginz2})
\begin{equation}
{d\sigma\over dy_1\,dy_2}=\left(\alpha\over\pi\right)^2R E^6\left(E\over M
\right)^8y_1^6f(y_1)y_2^6f(y_2),
\label{newdsigma}
\end{equation}
where, for spin 1/2,  (up to factors of order unity)
\begin{equation}
R\sim{10^{-4}\over\alpha^4}\left( n\over M\right)^8.
\end{equation}
The result in (\ref{newdsigma}) differs from that in Ref.~\cite{ginz2}
by a factor of $(E/M)^8y_1^4y_2^4$.  The photon distribution 
function $y^2f(y)$ used is rather strongly
peaked at $y\sim 0.3$.  (This peaking is necessary to have any chance
of satisfying the low-frequency criterion.)  When we multiply by $y^4$, the
amplitude is greatly reduced and the peak is shifted above $y=1/2$, violating
even the naive criterion for the validity of perturbation theory.  
Nevertheless, the integral of the distribution function is reduced by two 
orders of magnitude, that is,
\begin{equation}
{\int_0^1 dy\,y^6f(y)\over\int_0^1dy\,y^2 f(y)}\sim 10^{-2}.
\end{equation}
This reduces the mass limit quoted in \cite{d0} by a factor of $1/\sqrt{3}$,
to about 500 GeV, where $\langle\omega\rangle/M\approx 0.9$.  This
dubious result makes us conclude that
it is impossible to derive any limit for the monopole mass from the present
data.

As for the De R\'ujula limit\footnote{We note that De R\'ujula also considers
the monopole vacuum polarization correction to $g_V/g_A$, $g_A$, and $m_W/m_Z$,
proportional to $(m_Z/M)^2$ in each case, once again ignoring both the
string and the radiative correction problem.  He assumes that the
monopole is a heavy vector-like fermion, and obtains a limit of
$M/n>8m_Z$.  Our ansatz changes $(m_Z/M)^2$ to $(m_Z/M)^4$, so that
$M/\sqrt{n}>\sqrt{8}m_Z\approx250$ GeV, a substantial reduction.}
from the $Z\to3\gamma$ process, if we
insert a suppression factor of $\omega/M$ at each vertex and integrate over
the final state photon distributions, given by Eq.~(18) 
of Ref.~\cite{DeRujula},
the mass limit is reduced to $M/\sqrt{n}\gtrsim1.4m_Z\sim120$
GeV, again grossly violating the low
energy criterion. And the limit deduced from the vacuum polarization
correction to the anomalous magnetic moment of the muon due to virtual
monopole pairs \cite{graf} is reduced to 2 GeV.

The reader might object that this $\omega/M$ softening of the vertex has
little field-theoretic basis.  Therefore, we propose a second possibility that
does have such a basis.  The vertex (\ref{mmvertex}) suggests, and detailed
calculation supports (based on the tensor structure of the photon
amplitudes\footnote{For example, the naive monopole loop contribution to vacuum
polarization differs from that of an electron loop (apart from charge and
mass replacements) entirely by the replacement in the latter of
$(g_{\mu\nu}-q_\mu q_\nu/q^2)\to({\bf q}^2/q_0^2)(\delta_{ij}-q_iq_j/{\bf q}^2)
$, when $n^\mu$ points in the time direction. 
Apart from this different tensor structure, the vacuum polarization
is given by exactly the usual formula, found, for example in Ref.~\cite{js1}.
Details of this and related
calculations will be given in Ref.~\cite{gm}.}) the introduction of the 
string-dependent factor $\sqrt{q^2/(n\cdot q)^2}$ at each vertex, where $q$ is 
the photon momentum. Such a factor is devastating to the indirect monopole 
searches---for any process involving a real photon, such as that of the D0 
experiment \cite{d0} or for $Z\to3\gamma$ discussed in \cite{DeRujula}, the 
amplitude vanishes. Because such factors can and do appear in full monopole 
calculations, it is clearly premature to claim any limits based on virtual 
processes involving real final-state photons. 

\section{Conclusions}

We do not take our reduced limits on monopole masses very seriously.  Rather,
we believe they demonstrate our point that given the dual difficulties
of theoretically treating monopoles, that is, incorporating the string
and dealing with enormously strong coupling,\footnote{None of the papers
dealing with virtual monopole effects, \cite{DeRujula,ginz1,ginz2,graf},
in fact, incorporate these nontrivial effects.} it is premature to attempt
to set any limits on monopole masses based on virtual effects.  A direct
search stills seems much less problematic.

\section* {ACKNOWLEDGMENTS}

We thank Igor Solovtsov for helpful conversations and we are
grateful to the U.S.~Department of Energy for financial support.

\end{document}